\documentclass{amsproc}

\usepackage{amssymb}

\newtheorem{theorem}{Theorem}[section]

\newtheorem{proposition}[theorem]{Proposition}
\theoremstyle{definition}
\newtheorem{definition}[theorem]{Definition}

\theoremstyle{remark}

\numberwithin{equation}{section}

\DeclareMathOperator{\bv}{bv}
\DeclareMathOperator{\diag}{diag}
\newenvironment{psmallmatrix}
  {\left(\begin{smallmatrix}}
  {\end{smallmatrix}\right)}
\usepackage{color}

\begin{document}

\title[Sovable $n$-particle quantum graphs]{Solvable models of interacting 
$n$-particle systems on quantum graphs}

\author{Jens Bolte}
\address{Department of Mathematics, Royal Holloway, University of London, Egham, TW20 0EX, UK}
\curraddr{}
\email{jens.bolte@rhul.ac.uk}
\thanks{}

\author{George Garforth}
\address{Department of Mathematics, Royal Holloway, University of London, Egham, TW20 0EX, UK}
\curraddr{}
\email{George.Garforth.2012@live.rhul.ac.uk}
\thanks{}

\subjclass[2010]{81Q35}

\date{today}

\begin{abstract}
We introduce $n$-particle quantum graphs with singular two-particle interactions in 
such a way that eigenfunctions can be given in the form of a Bethe ansatz. We
show that this leads to a secular equation characterising eigenvalues of the Hamiltonian
that is based on a finite-dimensional determinant. These findings generalise previous
results about two-particle quantum graphs.
\end{abstract}

\maketitle

\section{Introduction}\label{Chapter6}

A quantum graph is metric graph equipped with a differential operator that
typically is a self-adjoint realisation of the differential Laplacian.
Spectral properties of quantum graphs recently attracted much attention (see, e.g.,
\cite{Exnetal08,BerKuc13}). They often mimic corresponding properties of Laplacians 
on manifolds and have become widely used models in quantum chaos \cite{KS99,GnuSmi06}. 
The popularity of quantum graph models is partly due to the fact that eigenvalues 
of compact quantum graphs can be characterised in terms of a secular equation 
involving a finite-dimensional determinant \cite{KS99}. This leads 
to a Gutzwiller-like trace formula that is an identity rather than an asymptotic 
relation \cite{KS99,KurNow05,BolEnd09}. Moreover, numerical calculations, with the aim of analysing the distribution of eigenvalues, can be performed 
very efficiently \cite{KS99}. 

More recently, many-particle quantum systems on graphs with singular interactions 
have been introduced \cite{BKContact,BKSingular}, primarily with the intention to 
study the existence or absence of Bose-Einstein condensation 
\cite{BolKerBEC,BolKerBEC2}. The latter mostly requires knowledge about low-lying 
eigenvalues, which can be gained without a secular equation. As the classical 
configuration space of $n\geq 2$~particles on a graph is $n$-dimensional, one 
would, in general, not expect to find a secular equation based on a finite-dimensional
determinant, but rather an approximate scheme involving a truncated determinant
as, e.g., developed in \cite{DorSmi92}. However, it is well known that particular 
many-particle quantum systems, such as the Lieb-Liniger gas on an interval or 
a circle \cite{LL63,Gau71}, are solvable by making use of an $n$-particle Bethe 
ansatz for the eigenfunctions \cite{Bet31}, which is a {\it finite} sum of plane waves.
In such a case only finitely many unknown coefficients need to be calculated in 
order to determine an eigenfunction. Although intervals and circles can be viewed 
as very simple graphs, an important step to generalising the Bethe ansatz to 
quantum graphs was done in \cite{CauCra07}, where the punctured real line, i.e.\ 
a graph with one vertex and two infinite edges, was equipped with a singular 
two-particle interaction that was specially designed to make a Bethe ansatz 
possible. In \cite{BolGar17} we extended this construction to two 
particles on arbitrary compact graphs, with explicit constructions for equilateral
star graphs as well as for tetrahedra. In all cases the Bethe ansatz led to a
secular equation with a finite-dimensional determinant. We numerically calculated
eigenvalues and analysed their distribution finding that, on the scale of the
mean level spacing, eigenvalues appear to be uncorrelated. As such an observation 
would normally be attributed to an integrable classical limit \cite{BT77}, we see 
it as a hint of an underlying quantum integrable field theory (as it
is the case for the Lieb-Liniger model). 

In what follows we generalise the approach to $n$-particle graphs. We maintain
the singular two-particle interactions between any pairs of the $n$ particles and 
show that the Bethe ansatz can be extended to this situation. We then derive a
secular equation, which is our main result.

\section{Preliminaries}

\subsection{One-particle quantum graphs}

We denote by $\Gamma$ a metric graph with finite vertex set $\mathcal{V}$ and
finite edge set $\mathcal{E}$. Each edge $e_j\in\mathcal{E}$ is assigned a (finite)
length $l_j$. The one particle Hilbert space is
\begin{equation}
L^2(\Gamma) = \bigoplus_{j=1}^{|\mathcal{E}|}L^2(0,l_j),
\end{equation}
and the Laplacian acting on $\Psi=(\psi_j)_{j=1}^{|\mathcal{E}|}\in H^2(\Gamma)$, with the Sobolev space $H^2(\Gamma)$ defined in analogy to $L^2(\Gamma)$, is
\begin{equation}
-\Delta_1\Psi = \left(-\psi_j''\right))_{j=1}^{|\mathcal{E}|}.
\end{equation}
Self-adjoint realisations of the Laplacian can be characterised in terms of
boundary conditions at the vertices \cite{KosSch99}. Let
\begin{align}
\Psi_{\mathrm{bv}}&=\left(\psi_1(0),\dots,\psi_{|\mathcal{E}|}(0),\psi_1(l_1),\dots,
                 \psi_{|\mathcal{E}|}(l_{|\mathcal{E}|})\right),\\
\Psi_{\mathrm{bv}}'&=\left(\psi_1'(0),\dots,\psi_{|\mathcal{E}|}'(0),-\psi_1'(l_1),\dots,
                 -\psi_{|\mathcal{E}|}'(l_{|\mathcal{E}|})\right),
\end{align}
be the vectors of boundary values and let $A,B$ be complex 
$|\mathcal{E}|\times|\mathcal{E}|$-matrices, such that $A B^\ast$ is self-adjoint
and $(A,B)$ has maximal rank. Then the boundary conditions
\begin{equation}\label{eq:AB}
A\Psi_{\mathrm{bv}}+B\Psi_{\mathrm{bv}}'=0
\end{equation} 
define a domain of a self-adjoint realisation of the Laplacian as a subspace of
$H^2(\Gamma)$. Further details can be found in, e.g., \cite{KosSch99,GnuSmi06,BerKuc13}.

The secular equation for Laplace eigenvalues requires two unitary 
$2|\mathcal{E}|\times 2|\mathcal{E}|$ matrices. One is the vertex $S$-matrix
\begin{equation}
S_v (k) = -(A+ikB)^{-1}(A-ikB),
\end{equation}
with $k\in\mathbb{R}$, and
\begin{equation}
T(k;\mathbf{l}) = \begin{pmatrix}  0 & e^{ik\mathbf{l}}\\ e^{ik\mathbf{l}} & 0\end{pmatrix},
\end{equation}
where
\begin{equation}
e^{ik\mathbf{l}} = \diag (e^{ikl_j})_{j=1}^{|\mathcal{E}|}.
\end{equation}
Then, $k^2>0$ is an eigenvalue of the Laplacian with boundary conditions \eqref{eq:AB}, iff
\begin{equation}\label{eq:1PSecular}
\det\left( \mathbb{I}-S_v (k) T(k;\mathbf{l}) \right) =0,
\end{equation}
see \cite{KS99,KosSch05}.

\subsection{Weyl group}

Before we proceed it is useful to define the symmetric group $S_n$ and the Weyl 
group $\mathcal{W}_n$, which we use to characterise the symmetries of exactly 
solvable $n$-particle systems.

Elements $Q$ of the symmetric group $S_n$ of order $n!$ will be written in terms 
of generators $T_1,\dots,T_{n-1}$ which satisfy the conditions
\begin{enumerate}
\item $T_iT_i=I$;
\item $T_iT_j=T_jT_i$ for $|i-j|>1$;
\item $T_iT_{i+1}T_i=T_{i+1}T_iT_{i+1}$.
\end{enumerate}
The $n-1$ generators of $S_n$ act on wave numbers $(k_1,\dots,k_n)$ according to
\begin{equation}\label{eq:Ti}
T_i(k_1,\dots,k_n)\equiv (k_{T_i(1)},\dots,k_{T_i(n)})
=(k_1,\dots,k_{i-1},k_{i+1},k_i,k_{i+2},\dots,k_n).
\end{equation}

\begin{definition}\label{def:Wn}
The Weyl group $\mathcal{W}_n$ is defined as a semidirect product of 
$(\mathbb{Z}/2\mathbb{Z})^n$ with the symmetric group $S_n$,
\begin{equation}
\mathcal{W}_n:=(\mathbb{Z}/2\mathbb{Z})^n\rtimes S_n .
\end{equation}
\end{definition}
Some useful and well-known properties of the Weyl group are the following:
\begin{proposition}\label{prop:W_n}
The Weyl group $\mathcal{W}_n$ has order $2^nn!$ and is generated by the elements
$T_1,\dots ,T_n,R_1$ that satisfy the conditions
\begin{enumerate}
\item $R_1R_1=I$;
\item $T_iT_i=I$;
\item $T_iT_j=T_jT_i$ for $|i-j|>1$;
\item $T_iT_{i+1}T_i=T_{i+1}T_iT_{i+1}$;
\item $R_1T_1R_1T_1=T_1R_1T_1R_1$;
\item $R_1T_i=T_iR_1$ for $i>1$.
\end{enumerate}
\end{proposition}
The $n$ generators of $\mathcal{W}_n$ act on wave numbers $(k_1,\dots,k_n)$ 
according to \eqref{eq:Ti} and 
\begin{equation}
R_1(k_1,\dots,k_n)=(-k_1,k_2,\dots,k_n).
\end{equation}
Finally it will be useful to relate the Weyl groups $\mathcal{W}_n$ and 
$\mathcal{W}_{n-1}$. To this end, it is convenient to define elements
\begin{align}
R_i=T_{i-1}\dots T_1R_1T_1\dots T_{i-1}
\end{align}
so that
\begin{align}
R_i(k_1,\dots,k_n)=(k_1,\dots,k_{i-1},-k_i,k_{i+1},\dots,k_n),
\end{align} 
and the cyclic permutation
\begin{align}
C_n=T_{n-1}T_{n-2}\dots T_1
\end{align}
so that
\begin{align}
C_n(k_1,k_2,\dots,k_{n-1},k_n)=(k_n,k_1,\dots,k_{n-2},k_{n-1}),
\end{align}
where we note the relation
\begin{align}\label{eq:Rn}
R_n=C_nR_1C_n^{-1}.
\end{align}
The Weyl group $\mathcal{W}_n$ can then be written in terms of $\mathcal{W}_{n-1}$ 
according to
\begin{align}\label{eq:Quotient}
\mathcal{W}_n=\left\{C_n^d(R_n)^j X;\ X\in\mathcal{W}_{n-1},\ d=0,\dots,n-1,\ j=0,1\right\}.
\end{align}
To simplify the notation we denote the action of any $P\in\mathcal{W}_n$ in analogy to 
\eqref{eq:Ti} as
\begin{equation}
P(k_1,\dots,k_n) = (k_{P(1)},\dots,k_{P(n)}).
\end{equation}

\subsection{Bosons in an interval}\label{sec:MultiESS}

In order to establish some key concepts in the $n$-particle setting, we begin by 
presenting the model of $n$ $\delta$-interacting bosons confined to an interval of 
length $l$ first solved by Gaudin \cite{Gau71}. The problem is formulated as a search 
for solutions of the formal Schr\"{o}dinger equation
\begin{align}\label{eq:MultinSE}
\left(-\Delta_n+2\alpha\sum_{i\neq j}\delta(x_i-x_j)\right)\psi(x_1,\dots,x_n)
=E\psi(x_1,\dots,x_n)
\end{align}
with $\alpha\in\mathbb{R}$ and particle positions $x_1,\dots,x_n$ on the half-line 
$\mathbb{R}_+=(0,\infty)$. Here the $n$-particle Laplacian acts according to 
\begin{align}\label{eq:MultiGFreeSchrodinger}
-\Delta_n\psi=-\sum_{j=1}^n\frac{\partial^2\psi}{\partial {x_j}^2}.
\end{align}

For a suitable choice of a domain, the Laplacian becomes
self-adjoint and provides a rigorous realisation of the formal 
operator in \eqref{eq:MultinSE}. Imposing bosonic symmetry
\begin{equation}
\psi(x_1,\dots,x_n)=\psi(x_{Q(1)},\dots,x_{Q(n)}),
\end{equation}
for all $Q\in S_n$,
this domain can be shown to consist of the set of $n-1$ jump 
conditions in the derivatives,
\begin{align}\label{eq:MultiLLCondition}
\left(\frac{\partial}{\partial x_{j+1}}-\frac{\partial}{\partial x_j}\right)\psi|_{x_{j+1}=x_j^+}=\alpha\psi|_{x_{j+1}=x_j^+},
\end{align}
for $j\in\{1,\dots,n-1\}$, and the Dirichlet condition
\begin{align}\label{eq:MultiDirichlet}
\psi|_{x_1=0}=0.
\end{align}
Due to the bosonic symmetry, $\psi$ can be restricted to the subspace
\begin{align}
d^I=\{(x_1,\dots,x_n)\in\mathbb{R}_+^n;\ x_1<\dots<x_n\}.
\end{align}
By applying a permutation $Q\in S_n$, it is also defined in all subspaces
\begin{align}
d^Q=\{(x_1,\dots,x_n)\in\mathbb{R}_+^n;\ x_{Q(1)}<\dots<x_{Q(n)}\},
\end{align}
and thus in all of $\mathbb{R}_+^n$. In the two-particle setting, interactions 
between particles means $\mathbb{R}_+^2$ is naturally dissected, along the line $x_1=x_2$, into two subspaces 
$d^I$ and $d^{T_1}$. Here, in the $n$-particle setting, appropriate dissections result in $n!$ 
subspaces labeled by elements $Q\in S_n$. Bosonic symmetry establishes equivalence 
between each of these subspaces so that we need only consider one.

The task is then to construct explicit Laplace eigenfunctions $\psi$ in $d^I$ 
which satisfy conditions \eqref{eq:MultiLLCondition} and \eqref{eq:MultiDirichlet}. 
The Bethe ansatz method in this context is the assumption that the appropriate 
ansatz is the sum of possible constituent plane wave states,
\begin{align}\label{eq:MultiGBethe}
\psi(x_1,\dots,x_n)=\sum_{P\in \mathcal{W}_n}\mathcal{A}^P e^{i(k_{P(1)}x_1+\dots+k_{P(n)}x_n)}.
\end{align}
A function of this form is an eigenfunction of the Laplacian with eigenvalue
\begin{align}\label{eq:MultiEigVals}
E=\sum_{j=1}^n k_j^2.
\end{align}
The $\delta$-type boundary conditions \eqref{eq:MultiLLCondition} which characterise 
interactions between particles, imply the relations
\begin{align}\label{eq:MultiGp}
\mathcal{A}^{PT_{i}}=s_p(k_{P(i)}-k_{P{(i+1)}})\mathcal{A}^P, 
\end{align}
for $i\in\{1,\dots,n-1\}$, with
\begin{align}
s_p(k)=\frac{k-i\alpha}{k+i\alpha}
\end{align}
for all $P\in\mathcal{W}_n$. The Dirichlet condition \eqref{eq:MultiDirichlet} at 
the end-point of the half line implies the relation
\begin{align}\label{eq:MultiGR}
\mathcal{A}^{PR}=-\mathcal{A}^P
\end{align}
for all $P\in\mathcal{W}_n$. Exact solvability is then assured if relations 
\eqref{eq:MultiGp} and \eqref{eq:MultiGR} are compatible with the properties of 
$\mathcal{W}_n$ as prescribed in Proposition~\ref{prop:W_n}. This amounts only 
to the requirement $s_p(u)s_p(-u)=1$ which is easily verified.

Until this point, particle positions have been defined on the half-line $\mathbb{R}^+$. Of course, in order to restrict the particles to an interval of length $l$ we must impose the further Dirichlet condition at the point $x=l$. Since bosonic symmetry allows us to restrict our attention to the domain $d^I\subset\mathbb{R}_+^n$, the appropriate boundary condition is given by
\begin{align}
\psi(x_1,\dots,x_{n-1},l)=0
\end{align}
which implies the relation
\begin{align}\label{eq:MultiRl}
\begin{split}
\mathcal{A}^P=&-e^{-2ik_{P(n)}l}\mathcal{A}^{PR_n}\\
=&-e^{-2ik_{P(n)}l}\mathcal{A}^{PC_nR_1C_n^{-1}},
\end{split}
\end{align}
where, for the latter equality, we have used \eqref{eq:Rn}. Finally, applying \eqref{eq:MultiGp}, \eqref{eq:MultiGR} and \eqref{eq:MultiRl} successively, we arrive at the condition
\begin{align}\label{eq:MultiGPSec}
e^{-2ik_{P(n)}l}=\prod_{i=1}^{n-1}s_p(k_{P(n)}+k_{P(i)})s_p(k_{P(n)}-k_{P(i)})
\end{align}
for all $P\in\mathcal{W}_n$. We note here that the form of $s_p(k)$ is such that, if \eqref{eq:MultiGPSec} is satisfied for  some $P\in\mathcal{W}_n$ then it is necessarily satisfied for elements
\begin{align}
PT_1,\dots, PT_{n-2},PR_1\text{ and }PR_{n}
\end{align}
in $\mathcal{W}_n$ and thus for every
\begin{align}
PX\text{ and }PR_nX
\end{align}
in $\mathcal{W}_n$ with $X\in\mathcal{W}_{n-1}$. Using \eqref{eq:Quotient}, we then have that the condition \eqref{eq:MultiGPSec} need only be satisfied for elements
\begin{align}
P\in\{I,C_n,C_n^2,\dots,C_n^{n-1}\}.
\end{align}
This is equivalent to the $n$ quantisation conditions
\begin{align}
e^{-2ik_jl}=\prod_{i\neq j}s_p(k_{j}+k_{i})s_p(k_{j}-k_{i}),
\end{align}
with $j\in\{1,\dots,n\}$. Solutions $(k_1,\dots,k_n)\neq(0,\dots,0)$, such that $0\leq k_1\dots \leq k_n$, then constitute energies \eqref{eq:MultiEigVals}.

\section{General graphs with $\tilde{\delta}$-interactions}

Now we have established how to construct exactly solvable $n$-particle systems on an interval, 
we would like to extend the approach to general graphs. It turns out that systems of $\delta$-interacting particles on graphs with more than a single edge, in general, are not compatibile with the Bethe ansatz method. In \cite{BolGar17}, which itself is based on a construction in 
\cite{CauCra07}, graphs were equipped with singular, non-local interactions, referred to as $\tilde{\delta}$-type, designed to make a Bethe ansatz possible. In what follows we extend this approach to $n$-particle quantum graphs. Defining 
an appropriate $n$-particle Bethe ansatz, we show exact solvability and calculate a quantisation 
condition, in the form of a collection of $n$ secular equations, which provide the exact spectra. 

Let us begin by viewing the compact graph $\Gamma$ in what we call its star representation 
$\Gamma^{(s)}$ by cutting all edges of $\Gamma$ to produce $|\mathcal{V}|$
star graphs and extending all the edges of the star graphs to infinity. The appropriate 
$n$-particle Hilbert space on $\Gamma^{(s)}$ is then
\begin{align}
\mathcal{H}_n^{(s)}=\bigotimes_{i=1}^{n}\left(\bigoplus_{j=1}^{|\mathcal{E}|} L^2(0,\infty)\right).
\end{align}
We remark that here $\mathcal{E}$ is now the union of the edge sets of all infinite star graphs.
Its order is twice the number of edges of the compact graph $\Gamma$. Vectors
\begin{align}
\Psi=\left(\psi_{j_1\dots j_n}^{(s)}\right)_{j_1,\dots,j_n=1}^{|\mathcal{E}|}
\end{align}
in $\mathcal{H}_n^{(s)}$ are then lists of $n$-particle functions
\begin{align}
\psi_{j_1\dots j_n}^{(s)}:{D}_{j_1\dots j_n}^{(s)}\rightarrow\mathbb{C}
\end{align}
in $L^2(D^{(s)}_{j_1\dots j_n})$ with infinite subdomains defined as
\begin{align}
{D}_{j_1\dots j_n}^{(s)}=(0,\infty)^n.
\end{align}
The total configuration space for $n$ particles on $\Gamma^{(s)}$ is the disjoint union
\begin{align}
D_{\Gamma}^{(s)}=\bigsqcup_{j_1,\dots,j_n=1}^{|\mathcal{E}|}D_{j_1\dots j_n}^{(s)}
\end{align}
of these subdomains. The $n$-particle Hilbert space can then be written $\mathcal{H}_n^{(s)}=L^2(D_{\Gamma}^{(s)})$. 

In the two-particle setting (see \cite{BolGar17}), interactions take place along the diagonals 
$x_1=x_2$ of two-dimensional configuration spaces ${D}_{mn}^{(s)}$. In the $n$-particle 
setting, we wish to impose interactions at the boundaries of subdomains
\begin{align}
D_{j_1\dots j_n}^{(s,Q)}=\{(x_1,\dots,x_n)\in D_{j_1\dots j_n}^{(s)};\ x_{Q(1)}<\dots <x_{Q(n)}\},
\end{align}
with $Q\in S_n$. The appropriate total dissected configuration space is then
\begin{align}\label{eq:MultiSRepDissection}
D_{\Gamma}^{(s,*)}=\bigsqcup_{j_1,\dots ,j_n=1}^{|\mathcal{E}|}\left(\bigsqcup_{Q\in S_n}{D}_{j_1\dots j_n}^{(s,Q)}\right),
\end{align}
with the total dissected two-particle Hilbert space 
$\mathcal{H}_n^{(s,*)}=L^2(D_{\Gamma}^{(s,*)})$. Thus vectors
\begin{align}
\Psi=
\left(\left(\psi_{j_1\dots j_n}^{(s,Q)}\right)_{j_1,\dots,j_n=1}^{|\mathcal{E}|}\right)_{Q\in S_n}
\end{align}
in $\mathcal{H}_n^{(s,*)}$ are lists of square-integrable functions 
$\psi_{j_1\dots j_n}^{(s,Q)}:D_{j_1\dots j_n}^{(s,Q)}\rightarrow \mathbb{C}$. The corresponding 
Sobolev space $H^2(D_{\Gamma}^{(s,*)})$ is the set of $\Psi\in\mathcal{H}_n^{(s,*)}$ consisting 
of functions $\psi_{j_1\dots j_n}^{(s,Q)}\in H^2(D_{j_1\dots j_n}^{(s,Q)})$. 

Boundary conditions will be imposed on eigenfunctions $\Psi\in H^2(D_{\Gamma}^{(s,*)})$ of the $n$-particle Laplacian $-\Delta_n$. We reiterate here that these will be $n$-particle analogues of the boundary conditions imposed in the two-particle setting in \cite{BolGar17}. Before we proceed with establishing these conditions, it is convenient to define the permutation matrices $\mathbb{Q}$ as representations of $Q\in S_n$ on
\begin{align}
\bigotimes_{j=1}^{n}\mathbb{C}^{|\mathcal{E}|}
\end{align}
such that
\begin{enumerate}
\item $\mathbb{I}=\mathbb{I}_{|\mathcal{E}|^n}$ is the representation of $I$;
\item $\mathbb{T}^{(i)}=\mathbb{I}_{|\mathcal{E}|^{i-1}}\otimes\mathbb{T}_{|\mathcal{E}|^2}\otimes\mathbb{I}_{|\mathcal{E}|^{n-i-1}}$ is the representation of $T_i$. 
\end{enumerate}
Here
\begin{equation}
\mathbb{T}_{|\mathcal{E}|^2}=\begin{pmatrix}  \mathbb{I}_{|\mathcal{E}|}\otimes m_1\\
\vdots\\ \mathbb{I}_{|\mathcal{E}|}\otimes m_{|\mathcal{E}|}\end{pmatrix},
\end{equation}
with the $|\mathcal{E}|$-dimensional row vectors
\begin{equation}
m_j = (0,\dots,0,1,0\dots,0)
\end{equation}
in which the $1$ is in $j$-th position. We note the properties 
\begin{align}
\mathbb{T}^{(i)}\left(\mathcal{A}_{j_1\dots j_n}\right)_{j_1,\dots, j_n=1}^{|\mathcal{E}|}
=\left(\mathcal{A}_{j_1\dots j_n}\right)_{j_1,\dots,j_{i-1},j_{i+1},j_{i},j_{i+2},\dots,j_n=1}^{|\mathcal{E}|}
\end{align}
for $|\mathcal{E}|^n$-dimensional column vectors $\mathcal{A}$ and that
\begin{align}
\begin{split}
\mathbb{T}^{(i)}(&M_1\otimes\dots \otimes M_{i-1}\otimes M_i\otimes M_{i+1}\otimes M_{i+2}\otimes\dots \otimes M_n)\mathbb{T}^{(i)}\\
=&M_1 \otimes\dots\otimes M_{i-1}\otimes M_{i+1}\otimes M_{i} \otimes M_{i+2}\otimes \dots \otimes M_n
\end{split}
\end{align}
for any $|\mathcal{E}|\times |\mathcal{E}|$ matrices $M_j$. Finally, it is convenient to note the property
\begin{align}\label{eq:KronPerm}
\mathbb{Q}\left(M\otimes\mathbb{I}_{|\mathcal{E}|^{n-1}}\right)\mathbb{Q}^{-1}=\mathbb{I}_{|\mathcal{E}|}\otimes\dots\otimes\mathbb{I}_{|\mathcal{E}|}\otimes M\otimes \mathbb{I}_{|\mathcal{E}|}\otimes\dots\otimes\mathbb{I}_{|\mathcal{E}|}
\end{align}
where on the right hand side, the matrix $M$ is the $Q(1)$-th position.

Let us begin by establishing boundary conditions which prescribe single-particle interactions 
with the vertices. These will be given as simple $n$-particle lifts of those given in \eqref{eq:AB} 
imposed in the one-particle setting. The values of $\Psi\in H^2(D_\Gamma^{(s,*)})$ at the 
vertices, along with corresponding derivatives, are given by boundary vectors
\begin{align}\begin{split}
\Psi_{\bv}^{(v)}&=\left((\psi_{j_1\dots j_n}^{Q}(x_1,\dots,x_n)|_{x_{Q(1)}=0})_{j_1,\dots,j_n=1}^{|\mathcal{E}|}\right)_{Q\in S_n};\\
{\Psi_{\bv}^{(v)}}'&=\left((\psi_{j_1\dots j_n,Q(1)}^{Q}(x_1,\dots,x_n)|_{x_{Q(1)}=0})_{j_1,\dots,j_n=1}^{|\mathcal{E}|}\right)_{Q\in S_n},
\end{split}
\end{align}
where $\psi_{j_1\dots j_n,Q(1)}^{Q}$ are inward derivatives normal to the lines $x_{Q(1)}=0$. Then, using \eqref{eq:KronPerm}, the appropriate boundary condition is given by
\begin{align}\label{eq:MultiVBV}
\left(\mathbb{I}_{n!}\otimes \mathbb{Q}\left(A\otimes \mathbb{I}_{|\mathcal{E}|^{n-1}}\right)\mathbb{Q}^{-1}\right)\Psi_{\bv}^{(v)}+\left(\mathbb{I}_{n!}\otimes \mathbb{Q}\left(B\otimes \mathbb{I}_{|\mathcal{E}|^{n-1}}\right)\mathbb{Q}^{-1}\right){\Psi_{\bv}^{(v)}}'=0.
\end{align}
Here the matrices $A,B$ are of the same form as in \eqref{eq:AB} .

Next, we would like to impose $\tilde{\delta}$-type interactions between pairs of particles located 
on the same infinite star and impose continuity across dissections otherwise. Such interactions 
are prescribed in the $n$-particle setting according to the conditions
\begin{align}\label{eq:MultiTildeCond}
\begin{split}
&\psi_{j_{Q^{-1}(1)}\dots j_{Q^{-1}(n)}}^Q(x_1,\dots,x_n)|_{x_{Q(i)}=x_{Q(i+1)}}\\
=&\psi_{j_{T_iQ^{-1}(1)}\dots j_{T_iQ^{-1}(n)}}^{QT_i}(x_1,\dots,x_n)|_{x_{Q(i)}=x_{Q(i+1)}};\\
&\left(\frac{\partial}{\partial x_{Q(i+1)}}-\frac{\partial}{\partial x_{Q(i)}}-2\alpha\right)\psi_{j_{Q^{-1}(1)}\dots j_{Q^{-1}(n)}}^Q(x_1,\dots,x_n)|_{x_{Q(i)}=x_{Q(i+1)}}\\
=&\left(\frac{\partial}{\partial x_{Q(i+1)}}-\frac{\partial}{\partial x_{Q(i)}}\right)\psi_{j_{T_iQ^{-1}(1)}\dots j_{T_iQ^{-1}(n)}}^{QT_i}(x_1,\dots,x_n)|_{x_{Q(i)}=x_{Q(i+1)}},
\end{split}
\end{align}
if the edges $e_{{j_i}}$ and $e_{j_{i+1}}$ belong to the same star graph, and
\begin{align}
\begin{split}
\label{eq:MultiContCond}
&\psi_{j_{Q^{-1}(1)}\dots j_{Q^{-1}(n)}}^Q(x_1,\dots,x_n)|_{x_{Q(i)}=x_{Q(i+1)}}\\
=&\psi_{j_{Q^{-1}(1)}\dots j_{Q^{-1}(n)}}^{QT_i}(x_1,\dots,x_n)|_{x_{Q(i)}=x_{Q(i+1)}};\\
&\left(\frac{\partial}{\partial x_{Q(i+1)}}-\frac{\partial}{\partial x_{Q(i)}}\right)\psi_{j_{Q^{-1}(1)}\dots j_{Q^{-1}(n)}}^Q(x_1,\dots,x_n)|_{x_{Q(i)}=x_{Q(i+1)}}\\
=&\left(\frac{\partial}{\partial x_{Q(i+1)}}-\frac{\partial}{\partial x_{Q(i)}}\right)\psi_{j_{Q^{-1}(1)}\dots j_{Q^{-1}(n)}}^{QT_i}(x_1,\dots,x_n)|_{x_{Q(i)}=x_{Q(i+1)}},
\end{split}
\end{align}
if the edges $e_{{j_i}}$ and $e_{j_{i+1}}$ belong to different star graphs.

The task is now to specify eigenvectors $\Psi\in H^2(D_\Gamma^{(s,*)})$ which satisfy boundary
conditions \eqref{eq:MultiVBV}-\eqref{eq:MultiContCond}. Taking care to distinguish between
subdomains $D_{j_1\dots j_n}^{(s,Q)}$, the vector $\Psi$ will be described by the collection 
of functions
\begin{align}
\psi^Q_{j_1\dots j_n}=\sum_{P\in\mathcal{W}_n}\mathcal{A}_{j_1\dots j_n}^{(P,Q)}e^{i(k_{P(1)}x_1+\dots+k_{P(n)}x_n)}.
\end{align}
This form obviously leads to eigenfunctions of $-\Delta_n$ with Laplace eigenvalues \eqref{eq:MultiEigVals}.

Let us define the $|\mathcal{E}|^n$-dimensional vectors
\begin{align}
\mathcal{A}^{(P,Q)}=\left(\mathcal{A}^{(P,Q)}_{j_1\dots j_n}\right)_{j_1,\dots,j_n=1}^{|\mathcal{E}|}
\end{align}
and then the $n!|\mathcal{E}|^n$-dimensional vectors
\begin{align}\label{eq:MultiAP}
\mathcal{A}^P=\left(\mathbb{Q}^{-1}\mathcal{A}^{(PQ^{-1},Q)}\right)_{Q\in S_n}.
\end{align}
It is convenient at this point to impose an ordering on \eqref{eq:MultiAP} by associating with each element $Q$ the number $[Q]\in (1,\dots,n!)$ so that
\begin{align}
\mathbb{Q}^{-1}\mathcal{A}^{(PQ^{-1},Q)} 
\end{align}
is the $[Q]\textsuperscript{th}$ block in the list $\mathcal{A}^P$.

Boundary conditions at the vertices \eqref{eq:MultiVBV} imply the relations
\begin{align}
\mathbb{Q}^{-1}\mathcal{A}^{(PR_{Q(1)},Q)}=\left(S_v(-k_{PQ(1)})\otimes\mathbb{I}_{|\mathcal{E}|^{n-1}}\right)\mathbb{Q}^{-1}\mathcal{A}^{(P,Q)}.
\end{align}
Noting then, that the properties of $\mathcal{W}_n$ imply
\begin{align}
R_{Q(1)}=QR_1Q^{-1},
\end{align}
we have that
\begin{align}\label{eq:MultiVertex}
\mathcal{A}^{PR_1}=\mathbb{I}_{n!}\otimes S_v(-k_{P(1)})\otimes\mathbb{I}_{|\mathcal{E}|^{n-1}}\mathcal{A}^P.
\end{align}

At this point, it is convenient to define the diagonal matrices
\begin{align}
\boldsymbol{c}_i=\diag(c^{(i)}_{j_1\dots j_n})_{{j_1,\dots,j_n=1}}^{|\mathcal{E}|},
\end{align}
where $c^{(i)}_{j_1\dots j_n}=1$, if the edges $e_{{j_i}}$ and $e_{j_{i+1}}$ belong to the same 
star graph, and $c^{(i)}_{j_1\dots j_n}=0$ otherwise. These matrices distinguish domains with 
$\tilde{\delta}$-type interactions from those which are continuous across dissections. We 
notice here the relations
\begin{align}
\boldsymbol{c}_i=\mathbb{I}_{|\mathcal{E}|^{i-1}}\otimes\boldsymbol{c}\otimes \mathbb{I}_{|\mathcal{E}|^{n-i-1}},
\end{align}
where $\boldsymbol{c}$ is defined for two-particle quantum graphs in \cite{BolGar17}. The $\tilde{\delta}$-type conditions \eqref{eq:MultiTildeCond} and continuity conditions \eqref{eq:MultiContCond} imply the relations
\begin{align}\label{eq:MultiCont}
\begin{split}
\left(\mathbb{I}_{2}\otimes \boldsymbol{c}_i\right)&\begin{pmatrix}
\mathbb{Q}^{-1}\mathcal{A}^{(PT_iQ^{-1},Q)}\\
\mathbb{T}^{(i)}\mathbb{Q}^{-1}\mathcal{A}^{(PQ^{-1},QT_i)}
\end{pmatrix}\\
=&\left(S_p(k_{P(i)}-k_{P(i+1)})\otimes\mathbb{I}_{|\mathcal{E}|^{n}}\right)\left(\mathbb{I}_{2}\otimes \boldsymbol{c}_i\right)
\begin{pmatrix}
\mathbb{Q}^{-1}\mathcal{A}^{(PQ^{-1},Q)}\\
\mathbb{T}^{(i)}\mathbb{Q}^{-1}\mathcal{A}^{(PT_iQ^{-1},QT_i)}
\end{pmatrix}
\end{split}
\end{align}
and
\begin{align}
(\mathbb{I}_{|\mathcal{E}|^n}-\boldsymbol{c}_i)\mathbb{Q}^{-1}\mathcal{A}^{(PQ^{-1},Q)}=(\mathbb{I}_{|\mathcal{E}|^n}-\boldsymbol{c}_i)\mathbb{Q}^{-1}\mathcal{A}^{(PQ^{-1},QT_i)}
\end{align}
respectively. We then have that
\begin{align}\label{eq:MultiT1}
\mathcal{A}^{PT_i}=Y_i(k_{P(i)}-k_{P(i+1)})\mathcal{A}^P,
\end{align}
where
\begin{align}\label{eq:MultiY1}
\left(Y_i(k)\right)_{[Q][Q']}=\frac{-i\alpha}{k+i\alpha}\boldsymbol{c}_i\delta_{[Q][Q']}+
\left(\frac{k}{k+i\alpha}\boldsymbol{c}_i+
\mathbb{T}^{(i)}\left(\mathbb{I}_{|\mathcal{E}|^n}-\boldsymbol{c}_i\right)\right)\delta_{[QT_i][Q']}.
\end{align}

Exact solvability is assured if relations \eqref{eq:MultiVertex} and \eqref{eq:MultiT1} are compatible with the properties of $\mathcal{W}_n$. This amounts to the following consistency relations:
\begin{enumerate}
\item $S_v(u)S_v(-u)=\mathbb{I}_{|\mathcal{E}|}$;
\item $Y_i(u)Y_i(-u)=\mathbb{I}_{n!|\mathcal{E}|^n}$;
\item $Y_i(u)Y_j(v)=Y_j(v)Y_i(u)$ for $|i-j|>1$;
\item $Y_{i+1}(u)Y_i(u+v)Y_{i+1}(v)=Y_i(v)Y_{i+1}(u+v)Y_i(u)$;
\item $\left(\mathbb{I}_{n!}\otimes S_v(u)\otimes\mathbb{I}_{|\mathcal{E}|^{n-1}}\right)Y_1(u+v)\left(\mathbb{I}_{n!}\otimes S_v(v)\otimes\mathbb{I}_{|\mathcal{E}|^{n-1}}\right)Y_1(v-u)\\
=Y_1(v-u)\left(\mathbb{I}_{n!}\otimes S_v(v)\otimes\mathbb{I}_{|\mathcal{E}|^{n-1}}\right)Y_1(u+v)\left(\mathbb{I}_{n!}\otimes S_v(u)\otimes\mathbb{I}_{|\mathcal{E}|^{n-1}}\right)$;
\item $Y_i(u)\left(\mathbb{I}_{n!}\otimes S_v(v)\otimes\mathbb{I}_{|\mathcal{E}|^{n-1}}\right)=\left(\mathbb{I}_{n!}\otimes S_v(v)\otimes\mathbb{I}_{|\mathcal{E}|^{n-1}}\right)Y_i(u)$ for $i>1$.
\end{enumerate}
These conditions can be verified by the explicit forms of $S_v(k)$ and $Y_i(k)$.

In order to turn the eigenfunctions in the star representation into eigenfunctions on the compact graph, we must impose appropriate joining conditions. These are the $n$-particle analogues of the joining conditions established in \cite{BolGar17} and are given by
\begin{align}
\psi_{j_1\dots j_n}^{Q}(x_1,\dots,x_n)=\psi_{j_1'\dots j_n'}^{Q}(x_1',\dots,x_n')
\end{align}
for all $Q\in S_n$, where
\begin{align}
(x_{Q(1)}',\dots,x_{Q(n)}')=(x_{Q(1)},\dots,x_{Q(n-1)},l_{j_{Q(n)}}-x_{Q(n)})
\end{align}
and
\begin{align}
(j_{Q(1)}',\dots,j_{Q(n)}')=(j_{Q(1)},\dots,j_{Q(n-1)},j_{Q(n)}+|\mathcal{E}|).
\end{align}
We then have
\begin{align}\label{eq:MultiJoining}
\begin{split}
\mathcal{A}^P=&E(-k_{P(n)})\mathcal{A}^{PR_n}\\
=&E(-k_{P(n)})\mathcal{A}^{PC_nR_1C_n^{-1}},
\end{split}
\end{align}
where
\begin{align}
E(k)=\mathbb{I}_{n!|\mathcal{E}|^{n-1}}\otimes
\begin{psmallmatrix}
0&1\\
1&0
\end{psmallmatrix}
\otimes
e^{ik\boldsymbol{l}}.
\end{align}

Applying \eqref{eq:MultiVertex}, \eqref{eq:MultiT1} and \eqref{eq:MultiJoining} successively we have that the relation
\begin{align}\label{eq:MultiEquiSec}
Z(k_{P(1)},\dots,k_{P(n)})=0,
\end{align}
with
\begin{align}\label{eq:MultiFinalEightQCon} 
\begin{split}
Z(k_1,\dots,k_n)=\det\Big[\mathbb{I}_{n!|\mathcal{E}|^n}-E(k_n)&Y_{n-1}(k_{n}-k_{n-1})\dots Y_1(k_{n}-k_{{1}})\\
\left(\mathbb{I}_{n!}\otimes S_v(k_{n})\otimes\mathbb{I}_{|\mathcal{E}|^{n-1}}\right)&Y_1(k_{1}+k_{n})\dots Y_{n-1}(k_{n-1}+k_{n})\Big],
\end{split}
\end{align}
is satisfied for all $P\in\mathcal{W}_n$. By using properties of determinants, it can be shown that the explicit forms of $Y_i(k)$, $S_v(k)$ and $E(k)$ are such that if \eqref{eq:MultiEquiSec} is satisfied for some $P\in\mathcal{W}_n$, then it is necessarily satisfied for elements
\begin{align}
PT_1,\dots ,PT_{n-2},R_1\text{ and }R_{n}
\end{align}
in $\mathcal{W}_n$ and thus for every
\begin{align}
PX\text{ and }PR_nX
\end{align}
in $\mathcal{W}_n$ with $X\in\mathcal{W}_{n-1}$. Using \eqref{eq:Quotient}, we can state the main result of this section.
\begin{theorem}\label{Theorem:MultiGeneralMainTheorem}
Non-zero eigenvalues of a self-adjoint $n$-particle Laplacian $-\Delta_n$ defined on $\Gamma$ with local vertex interactions specified through $A,B$ and $\tilde{\delta}$-type interactions between particles when they are located on neighbouring edges, are the values $E=k_1^2+\dots+k_n^2\neq 0$ with multiplicity $m$, where $(k_1,\dots,k_n)$, such that $0\leq k_1\leq\dots\leq k_n$, are solutions to the $n$ secular equations
\begin{align}\label{eq:MultiFinalQCon}
Z(k_{i_1},\dots,k_{i_n})=0,
\end{align}
for $(i_1,\dots,i_n)\in \{C_n^d(1,\dots,n)\}_{d=0}^{n-1}$, with multiplicity $m$.
\end{theorem}

\section{Recovering specific results}

In this final section,  by choosing particular parameters, we show how to recover established 
results from the general $n$-particle quantisation condition prescribed by Theorem \ref{Theorem:MultiGeneralMainTheorem}.

\subsection{Non-interacting particles}

Non-interacting systems are recovered by reestablishing continuity across all dissected domains. This is achieved by setting $\boldsymbol{c}_i=\boldsymbol{0}$ for all $i\in\{1,\dots,n-1\}$. Matrices 
\begin{align}
Y_i(k)|_{\boldsymbol{c}=\boldsymbol{0}}
\end{align}
are then composed of blocks
\begin{align}
\left(Y_i(k)|_{\boldsymbol{c}=\boldsymbol{0}}\right)_{[Q][Q']}=
\mathbb{T}^{(i)}\delta_{[QT_i][Q']}.
\end{align}
By substituting into \eqref{eq:MultiFinalEightQCon} we recover the secular equation 
\eqref{eq:1PSecular} for the one-particle quantum graph. 

\subsection{Two-particle graphs}

Simply by choosing $n=2$, one immediately recovers the spectra of two-particle quantum graphs as established in \cite{BolGar17} (see Theorem 5.1 therein).

\bibliographystyle{amsalpha}
\bibliography{Literature}

\newcommand{\etalchar}[1]{$^{#1}$}
\providecommand{\bysame}{\leavevmode\hbox to3em{\hrulefill}\thinspace}
\providecommand{\MR}{\relax\ifhmode\unskip\space\fi MR }
\providecommand{\MRhref}[2]{%
  \href{http://www.ams.org/mathscinet-getitem?mr=#1}{#2}
}
\providecommand{\href}[2]{#2}
\begin{thebibliography}{EKK{\etalchar{+}}08}

\bibitem[BE09]{BolEnd09}
J.~Bolte and S.~Endres, \emph{The trace formula for quantum graphs with general
  self-adjoint boundary conditions}, Ann. H. Poincar\'e \textbf{10} (2009),
  189--223.

\bibitem[Bet31]{Bet31}
H.~Bethe, \emph{Zur {T}heorie der {M}etalle. i. {E}igenwerte und
  {E}igenfunktionen der linearen {A}tomkette ({G}erman) [{O}n the theory of
  metals. i. {E}igenvalues and eigenfunctions of a linear chain of atoms]}, Z.
  Phys. \textbf{71} (1931), 205–226.

\bibitem[BG17]{BolGar17}
J.~Bolte and G.~Garforth, \emph{Exactly solvable interacting two-particle
  quantum graphs}, J. Phys. A: Math. Theor. (2017).

\bibitem[BK13a]{BerKuc13}
G.~Berkolaiko and P.~Kuchment, \emph{Introduction to quantum graphs},
  Mathematical Surveys and Monographs, vol. 186, AMS, 2013.

\bibitem[BK13b]{BKSingular}
J.~Bolte and J.~Kerner, \emph{Quantum graphs with singular two-particle
  interactions}, J. Phys. A: Math. Theor. \textbf{46} (2013), 045206.

\bibitem[BK13c]{BKContact}
\bysame, \emph{Quantum graphs with two-particle contact interactions}, J. Phys.
  A: Math. Theor. \textbf{46} (2013), 045207.

\bibitem[BK14]{BolKerBEC}
\bysame, \emph{Many-particle quantum graphs and {B}ose-{E}instein
  condensation}, J. Math. Phys. \textbf{55} (2014), 061901.

\bibitem[BK16]{BolKerBEC2}
J.~Bolte and J.~Kerner, \emph{Instability of {B}ose-{E}instein condensation
  into the one-particle ground state on quantum graphs under repulsive
  perturbations}, J. Math. Phys. \textbf{57} (2016), no.~4, 043301, 9.

\bibitem[BT77]{BT77}
M.~V. Berry and M.~Tabor, \emph{Level clustering in the regular spectrum},
  Proc. Roy. Soc. Ser. A \textbf{356} (1977), 375--394.

\bibitem[CC07]{CauCra07}
V.~Caudrelier and N.~Cramp\'e, \emph{Exact results for the one-dimensional
  many-body problem with contact interaction: including a tunable impurity},
  Rev. Math. Phys. \textbf{19} (2007), no.~4, 349--370.

\bibitem[DS92]{DorSmi92}
E.~Doron and U.~Smilansky, \emph{Semiclassical quantization of chaotic
  billiards: a scattering theory approach}, Nonlinearity \textbf{5} (1992),
  no.~5, 1055--1084.

\bibitem[EKK{\etalchar{+}}08]{Exnetal08}
P.~Exner, J.~P. Keating, P.~Kuchment, T.~Sunada, and A.~Teplyaev (eds.),
  \emph{Analysis on graphs and its applications}, Proc. Symp. Pure Math.,
  vol.~77, AMS, 2008, Papers from the program held in Cambridge, January
  8--June 29, 2007.

\bibitem[Gau71]{Gau71}
M.~Gaudin, \emph{Boundary energy of a {B}ose gas in one dimension}, Phys. Rev.
  A. \textbf{4} (1971), 386--394.

\bibitem[GS06]{GnuSmi06}
S.~Gnutzmann and U.~Smilansky, \emph{Quantum graphs: {A}pplications to quantum
  chaos and universal spectral statistics}, Advances in Physics \textbf{55}
  (2006), 527--625.

\bibitem[KN05]{KurNow05}
P.~Kurasov and M.~Nowaczyk, \emph{Inverse spectral problem for quantum graphs},
  J. Phys. A.: Math. Gen. \textbf{38} (2005), 4901–4915.

\bibitem[KS99a]{KosSch99}
V.~Kostrykin and R.~Schrader, \emph{Kirchhoff's rule for quantum wires}, J.
  Phys. A: Math. Gen. \textbf{32} (1999), 595--630.

\bibitem[KS99b]{KS99}
T.~Kottos and U.~Smilansky, \emph{Periodic orbit theory and spectral statistics
  for quantum graphs}, Ann. Physics \textbf{274} (1999), no.~1, 76--124.

\bibitem[KS06]{KosSch05}
V.~Kostrykin and R.~Schrader, \emph{Laplacians on metric graphs: Eigenvalues,
  resolvents and semigroups}, Contemp. Math. \textbf{415} (2006), 201--225.

\bibitem[LL63]{LL63}
E.~H. Lieb and W.~Liniger, \emph{Exact analysis of an interacting {B}ose gas.
  {I}. {T}he general solution and the ground state}, Phys. Rev. (2)
  \textbf{130} (1963), 1605--1616.

\end{thebibliography}

\end{document}